# SILICON MICRONEEDLES ARRAY WITH BIODEGRADABLE TIPS FOR TRANSDERMAL DRUG DELIVERY

*Bangtao Chen, Jiashen Wei, Francis E.H. Tay, Yee Ting Wong and Ciprian Iliescu*

Institute of Bioengineering and Nanotechnology, Singapore

## ABSTRACT

This paper presents the fabrication process, characterization results and basic functionality of silicon microneedles array with biodegradable tips. In order to avoid the main problems related to silicon microneedles: broking of the top part of the needles inside the skin, a simple solution can be fabrication of microneedles array with biodegradable tips. The silicon microneedles array was fabricated by using reactive ion etching while the biodegradable tips were performed using and anodization process that generates selectively porous silicon only on the top part of the skin. The paper presents also the results of in vitro release of calcein using microneedles array with biodegradable tips.

## 1. INTRODUCTION

Transdermal delivery is an attractive method to deliver drugs or biological compounds into human body, for its distinct advantage of eliminating pain and inconvenient intravenous injections. However, the efficiency of transdermal delivery is greatly limited by the poor permeability of the hard layer of skin at the stratum corneum which is the outmost layer of skin that forms the primary transport barrier [1]. The rate of diffusion also depends in part on the size and hydrophilicity of the drug molecules. So far, a number of chemical enhancers, electroporation, physical enhancers have been proposed to promote the transdermal drug delivery [2-4]. As one of the enhancers, the microneedle array devices have been well developed for controlled transdermal drug delivery in a minimum invasion and convenient manner [5-6]. The microneedles are used to penetrate the stratum corneum and generate pathways or microchannels, so to delivery drugs into the epidermis layer. No pain is induced as the needles do not reach the nerves in deep dermis.

Basically, the microneedles are fabricated by silicon with MEMS technologies. The various shapes and profiles of silicon microneedles, either hollow or solid needles, can be fabricated using the DRIE process [6-8]. However, due to the high-aspect-ratio needle structure and fragility of silicon, these microneedles have some several shortcomings. The main problem of the silicon microneedles is that it breaks easily during the insertion process into the skin (Figure 1), and this increases the possibility of an infection. The solution proposed here is to fabricate the microneedles tip from a biodegradable porous silicon material –. This porous silicon is well-known as a nano-structured silicon and is good for biological applications because of its bioactive and biodegradable properties [9-10]. The microneedles have been fabricated to have macro porous tips by using electrochemical etching process. These porous tips may break off after the drug delivery process and be allowed to remain in the skin, as it can be easily biodegraded within 2 to 3 weeks.

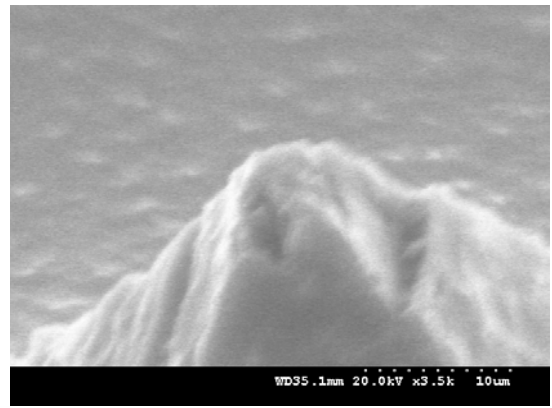

Figure 1. SEM image showing the tip of a broken microneedle after insertion into a pig skin.

Here we report the design and fabrication of microneedles array with biodegradable tips. The macroporous tips were fabricated using a classical anodization process in MeCN/HF/$H_2O$ solution. The microneedles array was tested using an *in vitro* animal skin model for the delivery of calcein. The drug release profile was characterized using UV spectrum detection, and the results showed that microneedles can enhance the drug release rate by 5 times compared to conventional transdermal drug delivery methods without enhancers.

## 2. DESIGN AND FABRICATION OF MICRONEEDLES

In order to penetrate the skin barrier of stratum corneum without reaching the nerves in the dermis layer, the microneedles should be at least 50 μm in length, but not more than 150 μm. Furthermore, to have an easy penetration into the skin, the microneedles should have a high aspect ratio. The fabrication of the microneedles was realized using an optimized $SF_6/O_2$ RIE process. By controlling the two gas flow rate, the microneedles were fabricated with an aspect ratio of 3:1 (height: width of the needle base). The isotropic profile of the microneedles was fabricated using two effects: flowing of thick photoresist mask and notching effect of the reflected charges on mask. It is a well-known property of the thick photoresist that it reflows when it is heated at 120°C [12]. The modification of the vertical wall of the patterned photoresist can reflect the charges (ions and radials) during the RIE process and in this way generates the etched profile under the oxide mask. This phenomenon has also been reported for the Bosch process [13]. Figure 2 is an example of the notching effect of reflected charges on the mask where ions and radicals with high energy are reflected by the oblique profile of the photoresist and generate an increased etching under the mask. The process eliminates the difficulty in the undercut control of the tips during the classical isotropic silicon etching process. Also, the end of the process can be very easily monitored using the video-camera of the "End Point Detection System" of the Deep RIE tool. Once the required shape of the tip is achieved, the square shape of the mask is modified and we could observe the different rectangular shapes that correspond to the projection of the tiled square mask on the monitor. An SEM image that illustrates the released needles at the masking layer is presented in Figure 3.

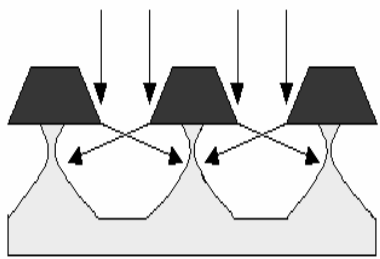

Figure 2. Notching effect of reflected charges on mask

The main steps of the fabrication process of the microneedles are presented in Figure 4. A 4" silicon wafer with <100> crystallographic orientation, n type 1-10 ohm-cm was initially cleaned in piranha solution ($H_2SO_4$: $H_2O_2$ 2:1) at 120°C for 20 minutes and then rinsed in DI water and spun dried (Figure 4a). On the silicon wafer a 0.5 μm-thick $SiO_2$ was deposited at 300°C, from $SiH_4$ and $N_2O$, at a pressure of 700 mTorr and a power of 300W using STS-PECVD equipment (Figure 4b). A photoresist mask using AZ9260 positive photoresist (from Clariant), with a thickness of 8 μm, was used for the patterning of the $SiO_2$-PECVD layer (Figure 4c). The pattern is transferred to the $SiO_2$ layer using a RIE etching system with $CHF_3$:He gases on RIE dielectrics Adixen AMS100 (Figure 4d). After the patterning of the oxide layer, the tips were generated using an isotropic RIE process (Figure 4e and 4f) with $SF_6/O_2$ in an ICP DRIE system. The process was optimized for a better control undercut, and with a depth-to-width aspect ratio of 3:1 [11]. A SEM image with the needle is presented in Figure 5.

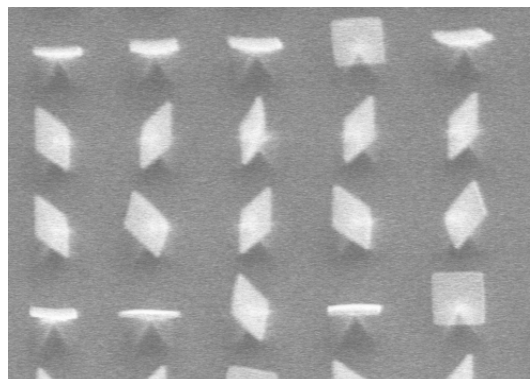

Figure 3: SEM image with released microneedles

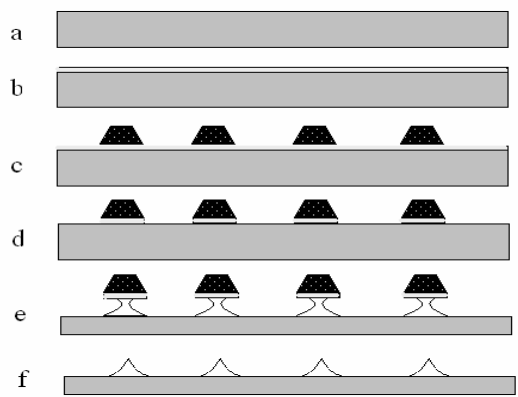

Figure 4. Fabrication process of nanotips: a) silicon wafer, b) PECVD $SiO_2$, c) Photoresist mask d) etching of $SiO_2$, e) plasma etching in $SF_6$, f) nanotips

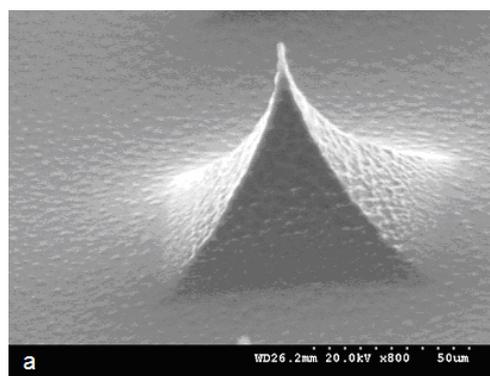

Figure 5. SEM picture of the fabricate microneedle

## 3. FABRICATION OF BIODEGRADABLE TIPS

After the fabrication of the silicon microneedles array, the next target was to develop an anodization process that allows the conversion of a single crystal silicon material of the tip into a porous structure. In order to achieve this, the "body" of the microneedles must be protected with a $Si_3N_4$ layer, releasing only the tip to be exposed during the anodization process. The main steps of the porous tips fabrication are shown in Figure 6.

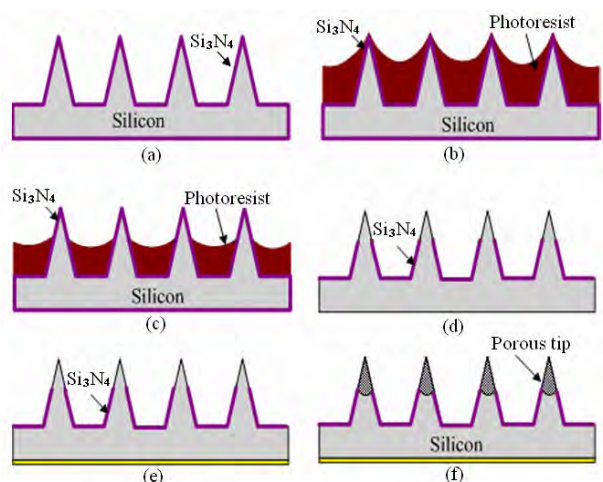

Figure 6. Fabrication process flow for the porous tips: (a) LPCVD $Si_3N_4$ deposition, (b) photoresist coating, (c) photoresist reflowing and $O_2$ plasma, (d) $Si_3N_4$ etching in RIE and photoresist removal, (e) Al deposition (back of the wafer), (f) anodization.

After cleaning, a 500 nm $Si_3N_4$ layer was deposited on the microneedles surface in LPCVD furnace at 725 °C (Figure 6a). A thick layer of photoresist AZ9260 was spun twice on the surface of the microneedles, followed by baking at 120°C for 15 min (Figure 6b). Due to the flowing effect, the photoresist covering the tips of the needles was much thinner than those at the body and at the bottom. Thus the photoresist on the tips of the microneedles was cleaned/removed in an $O_2$ plasma etching process in RIE mSystem (Figure 6c). In this way the top part of the needle is free of photoresist and the $Si_3N_4$ layer on the microneedle tips can be removed using a RIE process ($CHF_3$/He chemistry) -Figure 6d. The $Si_3N_4$ layer, from the backside of the wafer, was also removed using a similar RIE process. On the back of the wafer, an aluminum layer was sputtered in order to achieve a good electric contact for next electrochemical process (Figure 6e). Then the porous silicon was generated only on the tip of the microneedle using a classical anodization process, while as the needle body was protected by the remaining $Si_3N_4$ layer (Figure 6f).

Figure 7 shows the experimental setup of the anodic electrochemical etching process [14]. The Pt electrode was used as the cathode and silicon wafer as anode. A DC power of 36~72V was used as the source. The used electrolyte was a mixture of acetonitrile (MeCN) and diluted hydrofluoric acid (HF). The mixture has two compounds of MeCN:HF(4M):$H_2O$ = 92%:4%:4% by weight. The electrochemical anodization process was carried out at a current intensity of 10 mA $cm^{-2}$ for 30 min. Figure 8 shows a SEM picture of the porous silicon tip after the anodization process.

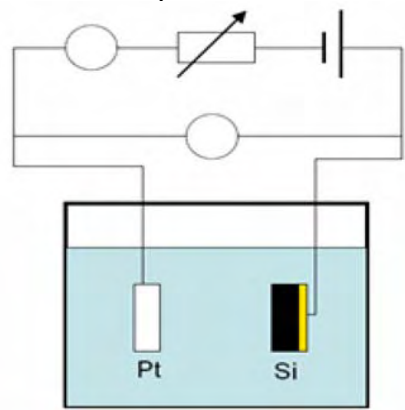

Figure 7. Schematic view of the electrochemical anodization process

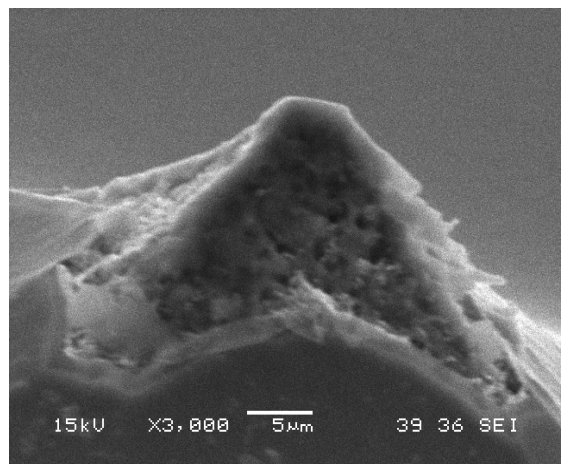

Figure 8. SEM picture of the porous silicon tip

## 4. CALCEIN DELIVERY USING MICRONEEDLES ARRAY WITH BIODEGRADABLE TIPS

The fabricated microneedles array was applied to the *in vitro* transdermal drug delivery model using animal skin tissues. Pig skins were used in the experiment due to their similar physiological properties with the human skin. All animal procedures were performed in compliance with relevant regulations approved by the Institutional Animal Care and Use Committee of National University of Singapore.

Microneedles were inserted into the pig skin to generate conduits or microchannels for the transport of drugs across the stratum corneum. Once the delivery compound crosses the stratum corneum, it diffuses rapidly through the deeper tissue and is taken up by the underlying capillaries for systemic administration. The transdermal drug delivery kinetics were studied using a calcein solution, where the permeability and transport of this chemical across the skin can be detected using the UV-visible spectrophotometric method.

For a better comparison, another *in vitro* test was performed without microneedles as the enhancer. The drug release profiles from the two kinds of transdermal delivery were compared and the results are shown in Figure 9. The skin permeability of calcein was greatly enhanced to 5 times with microneedles, compared to the transdermal delivery without microneedles.

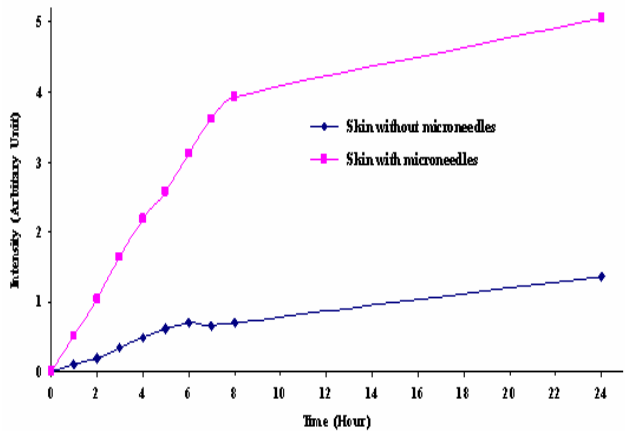

Figure 9. Calcein release profile with pig skin

## 5. CONCLUSIONS

The microneedles array with biodegradable porous silicon tips for transdermal drug delivery was fabricated using micromachining technologies. The high aspect ratio of these needles was obtained with reflow of photoresist and the notching RIE process. The microneedle tips fabricated are macroporous and biodegradable and this is achieved by using the electrochemical anodization process. The transdermal drug delivery experiments showed that the microneedles can greatly enhance the skin permeability for better drug transport. The results indicate the feasibility of microneedles as a more effective transdermal drug delivery system with significant clinical potential.

## ACKNOWLEDGEMENT

This project is funded by the Institute of Bioengineering and Nanotechnology, (IBN/04-R44007-OOE), Singapore.